\documentstyle[preprint,aps]{revtex}
\tighten
\begin{document}
\draft
\preprint{UTAPHY-HEP-13}
\title{Effective Mass Matrix for Light Neutrinos Consistent with
       Solar and Atmospheric Neutrino Experiments}
\author{S. P. Rosen and Waikwok Kwong}
\address{Department of Physics, University of Texas at Arlington,
         Arlington, Texas 76019-0059}
\date{January 17, 1995}
\maketitle

\begin{abstract}
We propose an effective mass matrix for light neutrinos which is
consistent with the mixing pattern indicated by solar and
atmospheric neutrino experiments. Two scenarios for the mass
eigenvalues are discussed and the connection with double beta decay
is noted.
\end{abstract}

\narrowtext
There are significant hints for neutrino mass coming from
solar\cite{solar} and atmospheric\cite{atmos} neutrino experiments
and they present an interesting theoretical challenge. Solar
neutrino experiments suggest that electron neutrinos oscillate into
other lepton flavors with a small mixing angle\cite{MSW}, $\sin^2
2\theta \approx 7\times 10^{-3}$, and a small mass-squared
difference, $\Delta m^2 \approx 5\times 10^{-6}~\rm eV^2$. Atmospheric
neutrino experiments, on the other hand, suggest that muon
neutrinos oscillate into tau neutrinos with maximal
mixing\cite{Fukuda}, $\sin^2 2\theta = 1$, and a much larger
mass-squared difference, $\Delta m^2 \approx 2\times 10^{-2}~\rm eV^2$.
How can we accommodated these very different sets of oscillation
parameters within one neutrino mass matrix?

Here we wish to propose an effective mass matrix, $\cal M$, of the
form:
\begin{eqnarray}
  \bar{\nu}{\cal M}\nu \equiv
  \pmatrix{ \bar{\nu}_e, & \bar{\nu}_\mu, & \bar{\nu}_\tau }
  \pmatrix{ m_1 &  m_2 &  m_2 \cr
            m_2 &  M_1 & -M_2 \cr
            m_2 & -M_2 &  M_1 \cr}
  \pmatrix{ \nu_e \cr \nu_{\mu} \cr \nu_{\tau} \cr}
\end{eqnarray}
which possesses the desired properties. The matrix is real and
symmetric and thus conserves CP irrespective of whether it
originates from Dirac or Majorana mass terms. Because the diagonal
elements in the $\mu$-$\tau$ sector are equal to one another, there
will be maximal mixing between these two flavors; furthermore, if
$m_2$ is taken to be much smaller than $M_1$ and $M_2$, then mixing
with the electron flavor will be weak. The same element $m_2$ is
used in the $e$-$\mu$ and $e$-$\tau$ positions so that matrix can be
diagonalised in a simple two-step process and the electron neutrino
can be decoupled from the heaviest mass eigenvector. The four
constants in the matrix can then be chosen to fit various scenarios
consistent with the oscillation hints. We shall consider two of
them below.

The first step in diagonalising the mass matrix is a rotation of
$45^\circ$ in the $\mu$-$\tau$ sector:
\begin{eqnarray}
U_1 & = & \pmatrix{1 &                0 &                 0 \cr
                   0 & \frac{1}{\sqrt2} & -\frac{1}{\sqrt2} \cr
                   0 & \frac{1}{\sqrt2} &  \frac{1}{\sqrt2} \cr},\\[0.5ex]
\tilde{U}_1{\cal M}U_1 & = &
          \pmatrix{       m_1 & \sqrt2 m_2 &          0 \cr
                   \sqrt2 m_2 &  M_1 - M_2 &          0 \cr
                            0 &          0 & M_1 + M_2  \cr}.
\end{eqnarray}
The second is to rotate the upper 2$\times$2 submatrix through an
angle $\theta$:
\begin{eqnarray}
U_2  = \pmatrix{ \cos \theta & -\sin \theta & 0 \cr
                 \sin \theta &  \cos \theta & 0 \cr
																											0 & 											0 & 1 \cr},
\end{eqnarray}
where
\begin{eqnarray}
\tan 2\theta = - \frac{2\sqrt2 m_2}{M_1 - M_2 - m_1}.
\end{eqnarray}
In the limit that $m_2$ is much smaller than the diagonal elements
of $\cal M$, the eigenvalues can be expressed in terms of a small
parameter $\Delta$,
\begin{eqnarray}
  M_x & = & m_1 - \Delta, \\
  M_y & = & M_1 - M_2 + \Delta, \\
  M_z & = & M_1 + M_2 \\[1ex]
  \Delta & = & \frac{2(m_2)^2}{M_1 - M_2 - m_1},
\end{eqnarray}
and
\begin{eqnarray}
  \tilde{U}_2 \tilde{U}_1 {\cal M} U_1 U_2 =
  \pmatrix{ M_x &   0 &   0 \cr
              0 & M_y &   0 \cr
              0 &   0 & M_z \cr}.
\end{eqnarray}

In terms of the mass eigenstates $\nu_w$ ($w = x, y, z$), the flavor
eigenstates become:
\begin{eqnarray}
  \pmatrix{ \nu_e \cr \nu_{\mu} \cr \nu_{\tau} \cr} & = &
  U_1U_2\pmatrix{ \nu_x \cr \nu_y \cr \nu_z \cr} \nonumber \\[0.5ex]
  & = & \pmatrix{ 1 & \frac{\Delta}{\sqrt2m_2} & 0 \cr
    -\frac{\Delta}{2m_2} & \frac{1}{\sqrt2} & - \frac{1}{\sqrt2} \cr
    -\frac{\Delta}{2m_2} & \frac{1}{\sqrt2} & + \frac{1}{\sqrt2} \cr}
  \pmatrix{ \nu_x \cr \nu_y \cr \nu_z \cr}
\end{eqnarray}
It follows from this expression that $\nu_e$ has no coupling to the
heaviest mass eigenvector $\nu_z$, and that it oscillates into a
coherent combination
\begin{eqnarray}
  \nu_a = \frac{1}{\sqrt2}(\nu_\mu + \nu_\tau)
\end{eqnarray}
of muon and tau neutrinos with mixing angle
\begin{eqnarray}
  \sin \theta_{ea} \approx \frac{\Delta}{\sqrt2 m_2} =
  \frac{\sqrt2 m_2}{M_1 - M_2 - m_1}.
\end{eqnarray}
The solar neutrino data\cite{solar,MSW}, $\sin^2 2\theta \approx
7\times 10^{-3}$, then implies that
\begin{eqnarray}
  \frac{\Delta}{\sqrt2 m_2} = \frac{\sqrt2 m_2}{M_1 - M_2 - m_1}
  \approx \frac{1}{23},
\end{eqnarray}
or
\begin{eqnarray}
  \frac{\Delta}{M_1 - M_2 - m_1} \approx \frac{1}{500}.
\end{eqnarray}

Having accommodated the mixing angles suggested by the solar and
atmospheric data, we now turn to the mass eigenvalues
$(M_x,M_y,M_z)$. The information available to us concerns
mass-squared differences, namely
\begin{eqnarray}
  \Delta_{yx} = (M_y)^2 - (M_x)^2 \approx 5\times 10^{-6}~\rm eV^2
\end{eqnarray}
from solar neutrino experiments\cite{solar,MSW}, and
\begin{eqnarray}
  \Delta_{zy} = (M_z)^2 - (M_y)^2 \approx 2\times 10^{-2}~\rm eV^2
\end{eqnarray}
from atmospheric ones\cite{Fukuda}. There is obviously a whole
family of solutions to these equations and in order to extract
interesting physics from them, we must make some assumption about
the  magnitudes of the masses in relation to the mass differences.
Two extreme possibilities are that the $(M_w)^2$ are either of the
same order as the $\Delta_{uv}$, or much greater than them.

In the former case, the parameters of the original mass matrix
$\cal M$ turn out to be:
\begin{eqnarray}
  m_1       &\ll&            10^{-3}~\rm eV, \nonumber\\
  m_2       & = &   7 \times 10^{-5}~\rm eV, \nonumber\\
  M_1 - M_2 & = &   2 \times 10^{-3}~\rm eV, \nonumber\\
  M_1 + M_2 & = & 1.4 \times 10^{-1}~\rm eV.
\end{eqnarray}
In the latter case, recent cosmological arguments\cite{Joel}
indicate masses of order 2 eV; taking this as our guide, we then
find that all three diagonal elements of $\cal M$ are close to 2
eV:
\begin{eqnarray}
  m_1       & = & 2~{\rm eV} - \delta - \eta, \nonumber\\
  m_2       & = & 4 \times 10^{-8}~\rm eV,    \nonumber\\
  M_1 - M_2 & = & 2~{\rm eV} - \delta,        \nonumber\\
  M_1 + M_2 & = & 2~{\rm eV} + \delta,        \nonumber\\[1ex]
  \delta    & = & 2.5 \times 10^{-3}~\rm eV,  \nonumber\\
  \eta      & = & 1.3 \times 10^{-6}~\rm eV.
\end{eqnarray}
It is interesting to note some numerical relationships between the
fine-tuned parameters of this form of $\cal M$. The parameter
$\delta$  is the square of $\frac{\Delta}{\sqrt2 m_2}$ and $\eta$ is
roughly the square of $\delta$; in addition $m_2$ is close to the
cube of $\delta$. This suggests that the mass matrix might be a power
series expansion of some relatively simple underlying matrix.

The closeness of $m_1$, the $e$-$e$ element of $\cal M$, to 2 eV
presents an interesting problem with regard to the Dirac versus
Majorana nature of the mass matrix. If $\cal M$ is constructed from
Majorana masses, then the amplitude for no-neutrino double beta
decay will be proportional to $m_1$. Now the bounds on the
effective Majorana mass for double beta decay are in the
neighborhood of 1--2 eV\cite{Moe}. Consequently we may be in a very
interesting position with respect to this lepton number violating
process: if the experimental sensitivity can be improved by an
order of magnitude\cite{Ge}, and if our $(M_w)^2 \gg \Delta_{uv}$
version of the mass matrix is correct, then we should actually
detect no-neutrino double beta decay.

Since $m_1$ is very small in the $(M_w)^2 \approx \Delta_{uv}$
version of $\cal M$, no-neutrino double beta decay would be
effectively undetectable and it cannot be determined whether this
version of the mass matrix is Majorana or Dirac. Thus a failure to
detect the process would indicate that either the $(M_w)^2 \approx
\Delta_{uv}$ mass matrix is correct, or that the $(M_w)^2 \gg
\Delta_{uv}$ one must be a Dirac mass matrix.

We have not constructed a specific gauge theory model for $\cal M$,
but we are confident that it is possible to do so. For example,
Albright and Nandi\cite{AN} have developed a procedure for starting
with low energy data and evolving SO(10) Grand Unified Theories
at the unification scale; indeed one of their scenarios matches the
mass eigenvalues of our $(M_w)^2 \approx \Delta_{uv}$ case and can
probably be adapted to yield a similar mixing matrix. Similarly,
Caldwell and Mohapatra\cite{CM}, motivated by cosmological
arguments, have considered a mass matrix similar to our $(M_w)^2
\gg \Delta_{uv}$ case in the context of SO(10) and left-right
symmetric models. Other models have been based upon radiative
corrections\cite{MP}, SUSY\cite{DHR}, and even
gravitation\cite{GS}, but the corresponding mass matrices tend to
have a different structure from the one proposed here.

This work was supported in part by the U.S. Department of Energy
Grant No. DE-FG05-92ER40691. One of the authors (SPR) is grateful
to the Aspen Center for Physics for its hospitality during the 1995
Winter Conference.

\end{document}